\documentclass[12pt]{iopart}

\usepackage{iopams}

\usepackage{graphicx}

\def\U#1{{%
\def\O{\mbox{O}}
\def\u{\mbox{u}}
\mathcode`\u=\mu
\mathcode`\O=\Omega
\mathrm{#1}}}

\def\ii{{\mathrm{i}}}

\def\ee{{\mathrm{e}}}

\def\dd{{\mathrm{d}}}

\def\Re{\mathop{\mathrm{Re}}}

\def\bra#1{\langle #1|}

\def\ket#1{|#1\rangle}

\def\bracket#1{\langle#1\rangle}

\def\bracketi#1#2{\langle#1|#2\rangle}

\def\bracketii#1#2#3{\langle#1|#2|#3\rangle}

\def\sub#1{_{\mathrm{#1}}}

\def\sur#1{^{\mathrm{#1}}}

\newcommand{\nn}{\nonumber \\}
\newcommand{\abs}[1]{\left\vert#1\right\vert}
\newcommand{\Real}{\mathbb{R}}

\DeclareMathSymbol{\varGamma}{\mathord}{letters}{"00}
\DeclareMathSymbol{\varDelta}{\mathord}{letters}{"01}
\DeclareMathSymbol{\varTheta}{\mathord}{letters}{"02}
\DeclareMathSymbol{\varLambda}{\mathord}{letters}{"03}
\DeclareMathSymbol{\varXi}{\mathord}{letters}{"04}
\DeclareMathSymbol{\varPi}{\mathord}{letters}{"05}
\DeclareMathSymbol{\varSigma}{\mathord}{letters}{"06}
\DeclareMathSymbol{\varUpsilon}{\mathord}{letters}{"07}
\DeclareMathSymbol{\varPhi}{\mathord}{letters}{"08}
\DeclareMathSymbol{\varPsi}{\mathord}{letters}{"09}
\DeclareMathSymbol{\varOmega}{\mathord}{letters}{"0A}

\begin{document}

\title{Geometrical aspects of weak measurements and quantum erasers}

\author{S Tamate$^1$, H Kobayashi$^1$, T Nakanishi$^{1,2}$, K Sugiyama$^{1,2}$, M Kitano$^{1,2}$}%

\address{$^1$ Department of Electronic Science and Engineering, Kyoto University, Kyoto 615-8510, Japan}

\address{$^2$ CREST, Japan Science and Technology Agency, Saitama 332-0012, Japan}

\ead{tamate@giga.kuee.kyoto-u.ac.jp}

\begin{abstract}
We investigate the mechanism of weak measurement by using an
interferometric framework.
In order to appropriately elucidate the interference effect that occurs
in weak measurement,
we introduce an interferometer for particles with internal degrees of freedom.
It serves as a framework common to quantum eraser and weak measurement.
We demonstrate that the geometric phase, particularly the Pancharatnam phase,
results from the post-selection of the internal state,
and thereby the interference pattern is changed.
It is revealed that the extraordinary displacement of the probe wavepackets
in weak measurement is achieved owing to the Pancharatnam phase
associated with post-selection.
\end{abstract}


\maketitle

\section{\label{sec:level1}Introduction}
Entanglement and interference are important phenomena in quantum mechanics,
and they sometimes lead to counterintuitive effects.
The theory of weak measurement, proposed by Aharonov, Albert, and Vaidman (AAV) \cite{aharonov_result_1988}, provides one of the most interesting examples
of such counterintuitive effects.
In weak measurements, the system state is post-selected after its
interaction with the probe system in addition to being pre-selected
in the state-preparation stage.
Moreover, the interaction is assumed to be weak, and therefore
the wavepackets of the probe remain overlapped.
Due to interference,
the average displacement in the position of the probe is proportional to
the real part of the so-called weak value \cite{aharonov_properties_1990},
\begin{equation}
 \bracket{\hat{A}}\sub{w} \equiv \frac{\bracketii{\psi\sub{f}}{\hat{A}}{\psi\sub{i}}}{\bracketi{\psi\sub{f}}{\psi\sub{i}}}, \label{eq:1}
\end{equation}
where $\hat{A}$ is a measured observable,
$\ket{\psi\sub{i}}$ is a pre-selected state,
and $\ket{\psi\sub{f}}$ is a post-selected state.
A possibly large displacement of the probe state according to the weak value
is called the AAV effect.
Shortly after the proposal of weak measurement,
an experimental scheme to observe the AAV effect was
presented by Duck {\it et al.}~\cite{duck_sense_1989}.
Thereafter, the AAV effect has been confirmed experimentally
using various optical systems
\cite{ritchie_realization_1991,suter_``weak_1995,parks_observation_1998,resch_experimental_2004,solli_fast_2004,brunner_direct_2004,pryde_measurement_2005,wang_experimental_2006,hosten_observation_2008,yokota_direct_2009,dixon_ultrasensitive_2009,kobayashi_simple_2009}.

Weak measurement provides a method
to measure the system state in a very weak interaction
that minimizes the disturbance to the system.
In fact, weak measurement is very useful for
experimentally detecting minute effects
\cite{hosten_observation_2008,dixon_ultrasensitive_2009},
because the weak value can lie outside the range of the eigenvalues of $\hat{A}$
for a small $\abs{\bracketi{\psi\sub{f}}{\psi\sub{i}}}$, as shown in Eq.~(\ref{eq:1}).
Its usefulness as a high-sensitivity measurement has
first been demonstrated by Hosten and Kwiat \cite{hosten_observation_2008}.
In their experiment, they measured the spin Hall effect of light
with a sensitivity of $0.1\,\U{nm}$.

Furthermore, weak measurement enables us to extract information
about quantum phases such as geometric phases \cite{sjoqvist_geometric_2006-1,shikano_weak_2008}.
It has been also shown that the weak value is closely related to
the phases of scattering matrices \cite{solli_fast_2004}.

The purpose of this study is to investigate the mechanism of weak measurement,
particularly phase changes at each stage, and thereby clarify
the physical meaning of the weak measurement.
For this purpose, we introduce an interferometer for particles
with internal degrees of freedom (spin or polarization).
It serves as a framework common to quantum eraser \cite{scully_quantum_1982}
and weak measurement.
In Sec.~\ref{sec:level2}, we first consider a quantum eraser from the aspect of
the phase change rather than the recovery of visibility due to the post-selection.
We demonstrate that the geometric phase \cite{berry_quantal_1984},
particularly the Pancharatnam phase \cite{pancharatnam_generalized_1956},
appears as a result of post-selection
in the quantum eraser.
In Sec.~\ref{sec:level3}, we examine the role of the post-selection in
the weak measurement.
We show that the extraordinary displacement of the probe wavepacket
in weak measurements is the result of a geometric property of the
Pancharatnam phase, which is induced by the post-selection.
The weak value can be geometrically understood in terms of
the behaviour of geodesic arcs on the Bloch (or Poincar\'e) sphere.

Recently, various applications based on weak measurement have been proposed and
experimentally demonstrated; for example, 
superluminal propagation \cite{solli_fast_2004,brunner_direct_2004},
entanglement concentration \cite{menzies_weak_2007},
and cross-phase modulation \cite{camacho_realization_2009}.
The geometric interpretation of the weak measurement will help us in
designing experimental schemes for such applications as well as
enable us to gain a comprehensive understanding of the weak measurement.

\section{\label{sec:level2}Pancharatnam phase in quantum erasers}
\begin{figure}
 \centering
 \includegraphics[width=68ex]{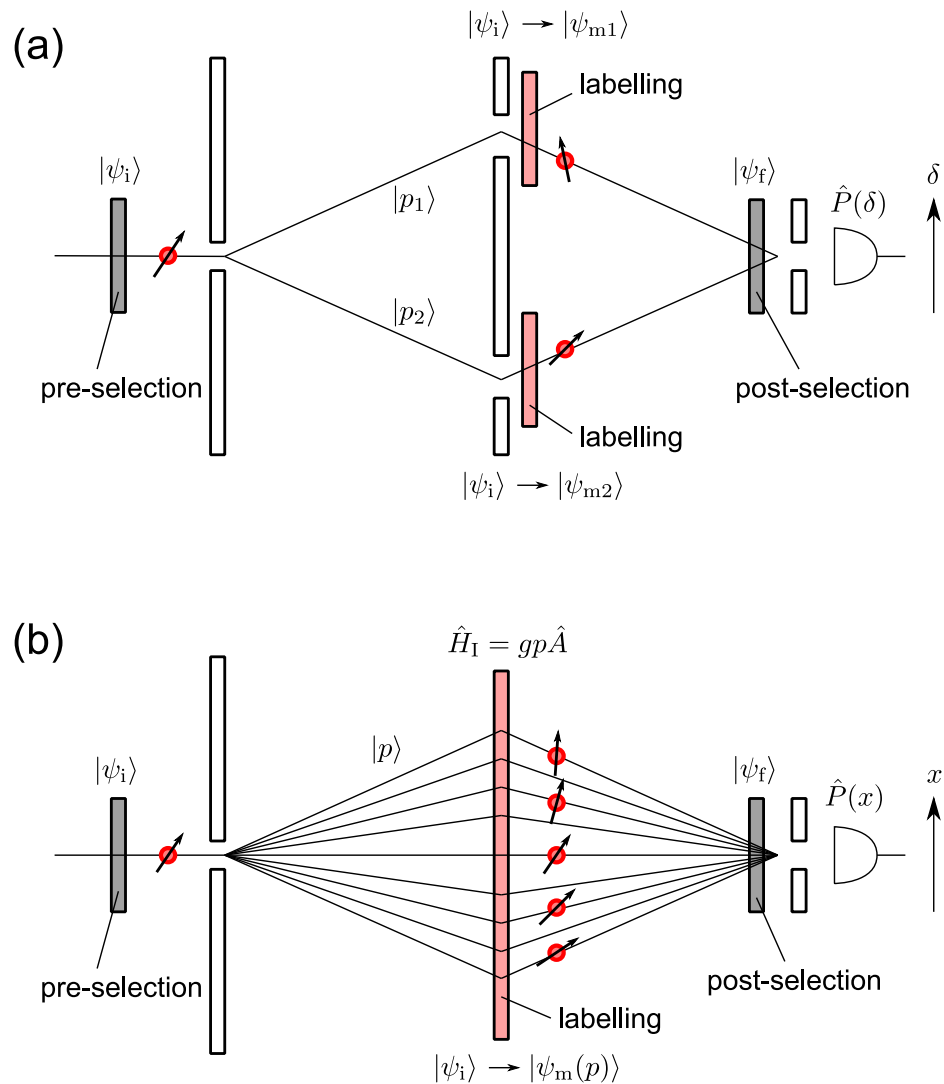}
 \caption{(a)~Experimental setup for quantum erasers.
 We label the paths by utilizing
 the particle's internal degree of freedom
 and erase the which-path information by post-selecting the internal state.
 (b)~Experimental setup for weak measurements.
 We label the momentum eigenstates by the Hamiltonian
 $\hat{H}\sub{I} = g\hat{A}\otimes \hat{p}$
 and post-select the internal state.}
 \label{fig:qe_wm}
\end{figure}

We consider a double-slit interferometer that can be used for a quantum particle,
as shown in Fig.~\ref{fig:qe_wm}~(a).
We assume that the particle has an internal degree of freedom.
In the quantum system, there exists a complementary relation between
which-path information and visibility of interference \cite{englert_fringe_1996}.
When we can extract the which-path information from the internal state,
the visibility of interference is decreased.
The idea of quantum eraser is that
one can erase the which-path information
by post-selecting the internal state, and then
the visibility of interference is recovered.
However, the post-selection of the internal state not only results in the recovery of visibility,
but also changes the phase of the interference.
In this section, we focus on the phase shift in the quantum eraser
and demonstrate that the phase shift induced by post-selection
can be expressed in terms of the Pancharatnam phase.

We set the initial state of the path as
\begin{equation}
 \ket{\phi\sub{i}} = c_1\ket{p_1} + c_2\ket{p_2},\quad c_1, c_2 \in \mathbb{C},\label{eq:2}
\end{equation}
where $\ket{p_1}$ and $\ket{p_2}$
correspond to the states of the upper and lower paths,
as shown in Fig.~\ref{fig:qe_wm}~(a).
We introduce the projection operator $\hat{P}(\delta)$ for determining
the relative phase of the paths as
\begin{equation}
 \hat{P}(\delta) = \ket{\phi(\delta)}\bra{\phi(\delta)}, \quad
 \ket{\phi(\delta)} = \frac{1}{\sqrt{2}}(\ket{p_1} + \ee^{\ii\delta}\ket{p_2}),
 \label{eq:3}
\end{equation}
and measure the interference pattern by sweeping the parameter $\delta$.
In order to calibrate the interferometer,
we first examine the initial interference pattern
and determine the phase $\delta\sub{i}$ that maximizes the detection probability,
\begin{equation}
 \Tr(\hat{P}(\delta)\ket{\phi\sub{i}}\bra{\phi\sub{i}})
  = \frac{1}{2}(\abs{c_1}^2 + \abs{c_2}^2 + c_1^*c_2\ee^{-\ii\delta} + c_1c_2^*\ee^{\ii\delta}). \label{eq:4}
\end{equation}
In this case, the phase is given by
\begin{equation}
 \delta\sub{i} = \arg(c_1^*c_2), \label{eq:5}
\end{equation}
which provides the origin of the phase,
and the choice of the origin depends on our calibration of the interferometer.

Secondly, we consider the internal degree of freedom
and assume that its initial state is $\ket{\psi\sub{i}}$.
The initial state of the joint system can be expressed as
\begin{equation}
 \ket{\varPsi\sub{i}} = \ket{\psi\sub{i}}\otimes(c_1\ket{p_1} + c_2\ket{p_2}).
  \label{eq:6}
\end{equation}
In order to label the particle according to the paths,
we let the initial state $\ket{\psi\sub{i}}$ evolve into the states
$\ket{\psi\sub{m1}}$ and $\ket{\psi\sub{m2}}$
corresponding to the paths $\ket{p_1}$ and $\ket{p_2}$, respectively.
Assuming that $\bracketi{\psi\sub{m1}}{\psi\sub{m2}} \neq 0$, we
cannot completely distinguish the paths.
Then, the state of the joint system can be expressed as the non-maximally entangled state,
\begin{equation}
 \ket{\varPsi\sub{m}} = c_1\ket{\psi\sub{m1}}\ket{p_1} + c_2\ket{\psi\sub{m2}}\ket{p_2}. \label{eq:7}
\end{equation}
The interference pattern is found to be
\begin{eqnarray}
 \Tr(\hat{P}(\delta)\ket{\varPsi\sub{m}}\bra{\varPsi\sub{m}})
 &= \frac{1}{2}(\abs{c_1}^2 + \abs{c_2}^2 \nn
 &\hspace{5ex} + c_1^*c_2\bracketi{\psi\sub{m1}}{\psi\sub{m2}}\ee^{-\ii\delta}
    + c_1c_2^*\bracketi{\psi\sub{m2}}{\psi\sub{m1}}\ee^{\ii\delta}), \label{eq:8}
\end{eqnarray}
and the phase $\delta\sub{m}$ that gives the maximum detection probability is
\begin{equation}
 \delta\sub{m} = \delta\sub{i} + \arg\bracketi{\psi\sub{m1}}{\psi\sub{m2}}. \label{eq:9}
\end{equation}
Thus, the phase shift $\delta^{(1)}$ due to the labelling is
\begin{equation}
\delta^{(1)} = \delta\sub{m} - \delta\sub{i} = \arg\bracketi{\psi\sub{m1}}{\psi\sub{m2}}. \label{eq:10}
\end{equation}
This implies that we can measure the intrinsic phase difference between
the internal states $\ket{\psi\sub{m1}}$ and $\ket{\psi\sub{m2}}$ as
the phase shift $\delta^{(1)}$.
The definition of the relative phase between two different states
as $\arg\bracketi{\psi\sub{m1}}{\psi\sub{m2}}$
was proposed by Pancharatnam \cite{pancharatnam_generalized_1956}.
When $\arg\bracketi{\psi\sub{m1}}{\psi\sub{m2}} = 0$ is satisfied,
$\ket{\psi\sub{m1}}$ and $\ket{\psi\sub{m2}}$ are known to be `in phase'.

Next, we examine the phase shift that is induced by post-selection
in the quantum eraser.
Due to the post-selection of the internal state in $\ket{\psi\sub{f}}$,
the state of the joint system becomes
\begin{eqnarray}
 \ket{\varPsi\sub{f}} &= \ket{\psi\sub{f}}\bra{\psi\sub{f}}\ket{\varPsi\sub{i}} \nn
 &= \ket{\psi\sub{f}}\otimes (c_1\bracketi{\psi\sub{f}}{\psi\sub{m1}}\ket{p_1} + c_2\bracketi{\psi\sub{f}}{\psi\sub{m2}}\ket{p_2}). \label{eq:11}
\end{eqnarray}
Then, constructive interference occurs at
\begin{eqnarray}
 \delta\sub{f} = \delta\sub{i} + \arg\bracketi{\psi\sub{m1}}{\psi\sub{f}}\bracketi{\psi\sub{f}}{\psi\sub{m2}}. \label{eq:12}
\end{eqnarray}
Hence, the phase shift $\delta^{(2)}$ that is induced by the post-selection is calculated as
\begin{eqnarray}
 \delta^{(2)} = \delta\sub{f} - \delta\sub{m} = \arg\bracketi{\psi\sub{m1}}{\psi\sub{f}}\bracketi{\psi\sub{f}}{\psi\sub{m2}}\bracketi{\psi\sub{m2}}{\psi\sub{m1}}. \label{eq:13}
\end{eqnarray}
This phase shift is gauge invariant; that is, it is independent of
the phase factor of each state.
Thus, the right hand side of Eq.~(\ref{eq:13}) represents the geometric phase,
particularly the so-called Pancharatnam phase
for the three states $\ket{\psi\sub{m1}}$, $\ket{\psi\sub{m2}}$ and $\ket{\psi\sub{f}}$ \cite{berry_adiabatic_1987,mukunda_quantum_1993}.

\begin{figure}
 \centering
 \includegraphics[width=50ex]{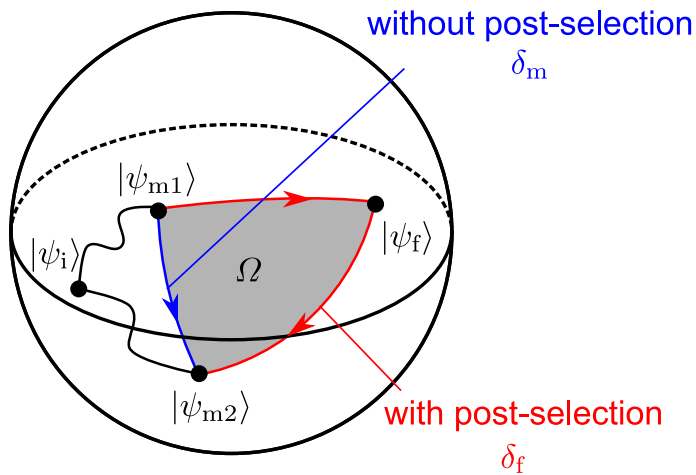}
 \caption{Pancharatnam phase on the Bloch sphere.
 The initial state $\ket{\psi\sub{i}}$ evolves into $\ket{\psi\sub{m1}}$
 and $\ket{\psi\sub{m2}}$; then, we compare the phases between them
 with or without post-selection in $\ket{\psi\sub{f}}$.
 The geodesic triangle formed by $\ket{\psi\sub{m1}}$, $\ket{\psi\sub{m2}}$,
 and $\ket{\psi\sub{f}}$ (shaded area) represents the Pancharatnam phase
 that results from post-selection.}
 \label{fig:bloch_evolution}
\end{figure}

Assuming that the particle has two internal states such as for polarization or spin $1/2$,
the Pancharatnam phase is known to be related to the solid angle $\varOmega$
(see Fig.~\ref{fig:bloch_evolution}) of the geodesic triangle
on the Bloch sphere by the following relation:
\begin{equation}
 \arg\bracketi{\psi\sub{m1}}{\psi\sub{f}}\bracketi{\psi\sub{f}}{\psi\sub{m2}}\bracketi{\psi\sub{m2}}{\psi\sub{m1}}
  = - \frac{\varOmega}{2}. \label{eq:14}
\end{equation}
Figure~\ref{fig:bloch_evolution} shows the relation between Eqs.~(\ref{eq:9})
and (\ref{eq:12}), each of which corresponds to interferometry {\it without}
and {\it with} post-selection, respectively.
In both procedures, the initial state $\ket{\psi\sub{i}}$ evolves into
$\ket{\psi\sub{m1}}$ and $\ket{\psi\sub{m2}}$ according to the corresponding paths,
and the phase difference between the two states is obtained by measuring
the interference pattern.
Without post-selection, we directly compare the phases between the two states
$\ket{\psi\sub{m1}}$ and $\ket{\psi\sub{m2}}$.
However, with post-selection, we compare the phases indirectly
via the post-selected state $\ket{\psi\sub{f}}$.
The difference between $\delta\sub{m}$ and $\delta\sub{f}$
is attributed to the Pancharatnam phase (\ref{eq:14})
and it can be obtained as the phase shift $\delta^{(2)} = \delta\sub{f} - \delta\sub{m}$.
The Pancharatnam phase for three states has been experimentally measured
using setups similar to that shown in Fig.~\ref{fig:qe_wm}~(a)
\cite{schmitzer_nonlinearity_1993,li_experimental_1999}.

We note that the phases $\delta\sub{i}$, $\delta\sub{m}$ and $\delta\sub{f}$
by themselves depend on our calibration of the interferometer.
On the contrast, the phase shifts $\delta^{(1)} = \delta\sub{m} - \delta\sub{i}$ and $\delta^{(2)} = \delta\sub{f} - \delta\sub{m}$ are
independent of the initial path state, and provide the phase information
about the internal state. The phase shift $\delta^{(1)}$ represents
the intrinsic phase difference between the two intermediate states
$\ket{\psi\sub{m1}}$ and $\ket{\psi\sub{m2}}$.
The phase shift $\delta^{(2)}$ represents the Pancharatnam phase
among the three states $\ket{\psi\sub{m1}}$, $\ket{\psi\sub{m2}}$
and $\ket{\psi\sub{f}}$, and critically depends on the choice of $\ket{\psi\sub{f}}$.

\section{\label{sec:level3}Reinterpretation of weak measurements}
\subsection{\label{sec:level3-1}The Pancharatnam-phase-induced displacement}
In this section, we describe how the Pancharatnam phase
contributes to the displacement of probe wavepackets in weak measurements
by applying the framework introduced in Sec.~\ref{sec:level2}.
Figure~\ref{fig:qe_wm}~(b) shows the experimental setup for the weak measurement.
We consider an interferometer that has many paths labelled with a continuous
variable $p$.
In this interferometer, the internal state of the particle
corresponds to the measured system state,
and the transverse (the $x$-direction) wavepacket corresponds to the probe state.
We assume the initial probe state to be
a Gaussian-like function centered at $p=0$ in the transverse momentum space.
Since we measure the position of the particle in weak measurements,
the analyzer operator $\hat{P}(x)$ is given by
\begin{equation}
 \hat{P}(x) = \ket{x}\bra{x}, \quad \ket{x} = \frac{1}{\sqrt{2\pi\hbar}} \int_\Real \ee^{-\ii xp/\hbar} \ket{p} \dd p,
  \label{eq:15}
\end{equation}
where $\ket{x}$ is the transverse position eigenstate
and $\ket{p}$ is the transverse momentum eigenstate.
The transverse position $x$ in Eq.~(\ref{eq:15}) determines the
phase gradient in the transverse momentum space
and plays the same role as the phase difference $\delta$ in Eq.~(\ref{eq:3}).
While we obtain the phase difference between the two paths
by measuring the constructive interference points in quantum eraser,
we obtain the phase gradient in the momentum space by measuring
the center of the wavepacket in weak measurement.
The phase calibration of the interferometer in quantum eraser
corresponds to the prior determination of the center of the wavepacket
in weak measurement.

We label each momentum eigenstate under
the interaction Hamiltonian $\hat{H}\sub{I} = g\hat{A}\otimes\hat{p}$,
where $g$ is the coupling constant
and $\hat{p}$ is the transverse momentum operator.
After the interaction for a time period $\tau$,
$\ket{\psi\sub{i}}$ evolves into
$\ket{\psi\sub{m}(p)}=\ee^{-\ii G p \hat{A} / \hbar}\ket{\psi\sub{i}}$
according to the path $\ket{p}$, where $G = g\tau$.
This process leads to the phase difference
$\varTheta\sur{(1)}(p)$ between the momentum eigenstates
$\ket{p=0}$ and $\ket{p}$:
\begin{eqnarray}
 \varTheta\sur{(1)}(p) &= \arg\bracketi{\psi\sub{m}(0)}{\psi\sub{m}(p)}
 =\arg\bracketii{\psi\sub{i}}{\ee^{-\ii G p \hat{A} / \hbar}}{\psi\sub{i}}
 \sim  - \frac{G\bracket{\hat{A}}}{\hbar}p. \label{eq:16}
\end{eqnarray}
The phase change $\varTheta^{(1)}(p)$ can be regarded as
the dynamical phase \cite{berry_quantal_1984},
which is proportional to the energy of the particle.
In fact, $\varTheta^{(1)}$ is expressed as
\begin{equation}
\varTheta^{(1)}(p) \sim
-\frac{\bra{\psi\sub{i}}\bracketii{p}{\hat{H}\sub{I}}{\psi\sub{i}}\ket{p}\tau}{\hbar}
= -\frac{G\bracket{\hat{A}}}{\hbar}p. \label{eq:17}
\end{equation}
The $p$-dependent phase shift changes the constructive interference point
and is measured as the displacement of the wavepacket.
The displacement $\Delta x^{(1)}$
due to the labelling is given by
\begin{equation}
 \Delta x^{(1)} = -\hbar\left.\frac{\dd \varTheta^{(1)}}{\dd p}\right|_{p = 0}
 = G\bracket{\hat{A}}. \label{eq:18}
\end{equation}
Thus, we can obtain the expectation value of the observable $\hat{A}$.

In addition, as shown in Eq.~(\ref{eq:13}),
when we post-select the internal state in $\ket{\psi\sub{f}}$,
the Pancharatnam phase $\varTheta^{(2)}(p)$ appears as an additional phase shift:
\begin{eqnarray}
 \varTheta^{(2)}(p) &=
  \arg \bracketi{\psi\sub{m}(0)}{\psi\sub{f}}
  \bracketi{\psi\sub{f}}{\psi\sub{m}(p)}
  \bracketi{\psi\sub{m}(p)}{\psi\sub{m}(0)} \nn
  &= \arg \left[
  \bracketi{\psi\sub{i}}{\psi\sub{f}}
  \bra{\psi\sub{f}}\ee^{- \ii G p \hat{A} / \hbar}\ket{\psi\sub{i}}
  \bra{\psi\sub{i}}\ee^{ \ii G p \hat{A} / \hbar}\ket{\psi\sub{i}}
  \right] \nn
  &\sim -\frac{G(\Re\bracket{\hat{A}}\sub{w} - \bracket{\hat{A}})}{\hbar}p. \label{eq:19}
\end{eqnarray}
Hence, the displacement $\Delta x^{(2)}$ caused by the post-selection is
\begin{equation}
 \Delta x^{(2)} = - \hbar \left.\frac{\dd \varTheta^{(2)}}{\dd p}\right|_{p = 0}
  = G (\Re \bracket{\hat{A}}\sub{w} - \bracket{\hat{A}}). \label{eq:20}
\end{equation}

After all, the displacement $\Delta x$ for the whole process of weak measurement is the sum of
$\Delta x^{(1)}$ and $\Delta x^{(2)}$:
\begin{equation}
 \Delta x = \Delta x^{(1)} + \Delta x^{(2)} = G\Re\bracket{\hat{A}}\sub{w}. \label{eq:21}
\end{equation}
Consequently, the displacement $\Delta x$ is
obtained as the real part of the weak value $\bracket{\hat{A}}\sub{w}$.
The counterintuitive effects in weak measurement such as
the unbounded weak value can be attributed to
the Pancharatnam-phase-induced displacement $\Delta x^{(2)}$,
as will be shown in Sec.~\ref{sec:level3-2}.

\subsection{\label{sec:level3-2}Phase jump in the Pancharatnam phase}
In weak measurements, the smaller the inner product of $\ket{\psi\sub{i}}$ and
$\ket{\psi\sub{f}}$, the larger is the displacement $\Delta x$,
as shown in Eqs.~(\ref{eq:1}) and (\ref{eq:21}).
This effect is closely related to the phase jump in the Pancharatnam phase
that is caused by the geometrical singularity of geodesics on the Bloch sphere
\cite{schmitzer_nonlinearity_1993,li_experimental_1999,bhandari_su_1991}.
As an example, we consider a two-state system as a measured system and 
denote its basis states by $\ket{+}$ and $\ket{-}$.
The initial state  $\ket{\psi\sub{i}}$, the post-selected state $\ket{\psi\sub{f}}$,
and the observable $\hat{A}$ are defined as follows:
\begin{eqnarray}
 \ket{\psi\sub{i}} &= \ket{+}, \label{eq:22}\\
 \ket{\psi\sub{f}} &= \sin\theta\ket{+} + \cos\theta\ket{-}, \label{eq:23}\\
 \hat{A} &= \ket{+}\bra{-} + \ket{-}\bra{+}. \label{eq:24}
\end{eqnarray}
The expectation value and the weak value of $\hat{A}$ are
$\bracket{\hat{A}} = \bracketii{\psi\sub{i}}{\hat{A}}{\psi\sub{i}} = 0$
and $\bracket{\hat{A}}\sub{w}
= \bracketii{\psi\sub{f}}{\hat{A}}{\psi\sub{i}}/\bracketi{\psi\sub{f}}{\psi\sub{i}} = 1/\tan\theta$, respectively.
The system state $\ket{\psi\sub{m}(p)}$ that is evolved corresponding
to the probe state $\ket{p}$ is given by
\begin{equation}
 \ket{\psi\sub{m}(p)} = \ee^{-\ii G p \hat{A} / \hbar}\ket{\psi\sub{i}}
 = \cos\varphi\ket{+}
  - \ii \sin\varphi\ket{-}, \label{eq:25}
\end{equation}
where $\varphi(p) = Gp/\hbar$.
The additional phase shift induced between
the momentum eigenstates $\ket{p=0}$ and $\ket{p}$
by post-selection is derived as
\begin{equation}
 \varTheta^{(2)}(p) = \arg\bracketi{\psi\sub{i}}{\psi\sub{f}}
 \bracketi{\psi\sub{f}}{\psi\sub{m}(p)}
 \bracketi{\psi\sub{m}(p)}{\psi\sub{i}}
 = - \tan^{-1}\left(\frac{\tan\varphi}{\tan\theta}\right)\!. \label{eq:26}
\end{equation}

\begin{figure}
 \centering
 \includegraphics[width=50ex]{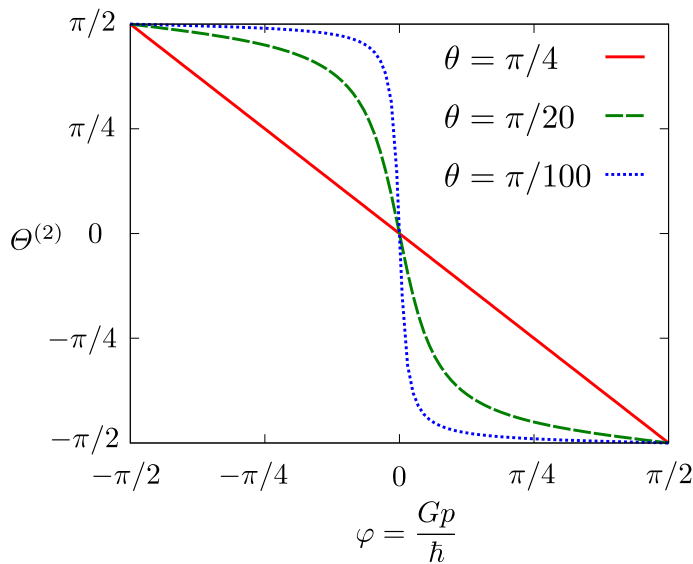}
 \caption{Variation of Pancharatnam phase $\varTheta^{(2)}$ for several $\theta$.
 The gradient of the Pancharatnam phase becomes steeper with decreasing $\theta$.
 Since the Pancharatnam phase obtained around $p=0$ is limited to
 $\pi$, the region in which the Pancharatnam phase changes linearly
 becomes smaller for the smaller $\theta$.
 }
 \label{fig:plot}
\end{figure}

\begin{figure}
 \centering
 \includegraphics[width=40ex]{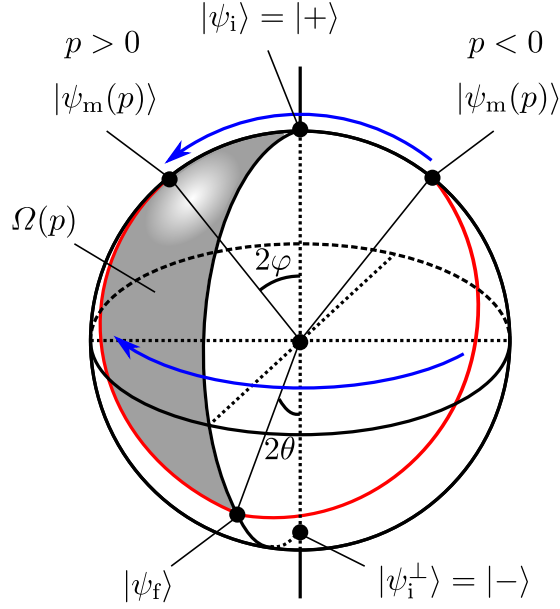}
\caption{Variation of geodesic triangle on Bloch sphere.
 The initial internal state $\ket{\psi\sub{i}}$ corresponds to the north pole 
 $\ket{+}$, and the post-selected state $\ket{\psi\sub{f}}$
 occurs near the south pole $\ket{-}$.
 After the interaction, the internal state is rotated clock-wise
 or anti-clockwise into $\ket{\psi\sub{m}(p)}$ according to $p < 0$ or $p >0$.
 By the post-selection, the transverse momentum eigenstate $\ket{p}$ acquires
 the Pancharatnam phase $\varTheta^{(2)}(p) = -\varOmega(p)/2$.
 When $\ket{\psi\sub{m}(p)}$ traverses the north pole $\ket{+}$,
 the geodesic arc connecting $\ket{\psi\sub{m}(p)}$ and $\ket{\psi\sub{f}}$
 rapidly sweeps across the surface of the Bloch sphere,
 and therefore the Pancharatnam phase also changes rapidly
 around $p=0$.
}
 \label{fig:bloch}
\end{figure}

We show the variation in $\varTheta^{(2)}$ for several post-selected states
in Fig.~\ref{fig:plot}.
The trend in the phase change can be well understood by considering
the geometrical meaning of the Pancharatnam phase.
Figure~\ref{fig:bloch} shows the variation of the geodesic triangle on the Bloch sphere.
The initial state $\ket{\psi\sub{i}}$ corresponds to the north pole $\ket{+}$
and the post-selected state $\ket{\psi\sub{f}}$ occurs near the south pole $\ket{-}$.
The solid angle $\varOmega(p)$ of the geodesic triangle connecting
$\ket{\psi\sub{i}}$, $\ket{\psi\sub{m}(p)}$, and $\ket{\psi\sub{f}}$
is related to the Pancharatnam phase by the relation
$\varTheta^{(2)}(p) = -\varOmega(p)/2$ as shown in Eq.~(\ref{eq:14}).

For simplicity, we assume $0 < \theta \ll \pi/4$ and set $\ket{\psi\sub{i}^\bot} = \ket{-}$.
We sweep $\varphi$ for a fixed value of $\theta$.
For $\varphi > \theta$, the distance between
$\ket{\psi\sub{m}(p)}$ and $\ket{\psi\sub{i}}$ becomes large as compared to
that between $\ket{\psi\sub{f}}$ and $\ket{\psi\sub{i}^\bot}$.
Therefore, the path of the geodesic arc connecting 
$\ket{\psi\sub{m}(p)}$ and $\ket{\psi\sub{f}}$
passes close to the path connecting $\ket{\psi\sub{m}(p)}$ and $\ket{\psi\sub{i}^\bot}$.
Since, in this example, the geodesic arc connecting
$\ket{\psi\sub{m}(p)}$ and $\ket{\psi\sub{i}^\bot}$
is always on the same great circle,
the geodesic arc connecting
$\ket{\psi\sub{m}(p)}$ and $\ket{\psi\sub{f}}$ remains almost constant.
As a result, the variation in the Pancharatnam phase
in the range $\varphi > \theta$ is quite small.
This is also true in the range $\varphi < -\theta$.
For $\varphi < 0$, however, the geodesic arc connecting
$\ket{\psi\sub{m}(p)}$ and $\ket{\psi\sub{i}^\bot}$
goes in the opposite direction around the Bloch sphere,
as compared to that in the case of $\varphi > 0$.
Thus, the geodesic arc connecting $\ket{\psi\sub{m}(p)}$ and $\ket{\psi\sub{f}}$
must change rapidly in the range $-\theta < \varphi < \theta$.
This is why the Pancharatnam phase jumps by $\pi$ around $p = 0$.
As shown in Fig.~\ref{fig:plot}, the smaller the value of $\theta$,
the steeper is the gradient of $\varTheta^{(2)}(p)$.

Weak measurements utilize the large gradient of the Pancharatnam phase
around $p = 0$.
Since $\bracket{\hat{A}} = 0$ in this example, the real part of 
the weak value is proportional to the gradient of the Pancharatnam phase:
\begin{equation}
 \Delta x = G\Re\bracket{\hat{A}}\sub{w} =
  - \hbar\left.\frac{\dd \varTheta^{(2)}}{\dd p}\right|_{p=0}.
\end{equation}
Therefore, when $\bracketi{\psi\sub{f}}{\psi\sub{i}} \sim \theta$ is small,
we can obtain the large displacement.

The Pancharatnam phase varies nonlinearly with $p$;
therefore, in order to maintain the shape of the wavepacket,
the momentum distribution of the wavepacket must be contained in the
range in which the Pancharatnam phase changes linearly \cite{kitano_negative_2003}.
Let $\Delta p$ be the momentum variance,
then the condition under which the Pancharatnam phase varies linearly
is given by $\varphi(\Delta p) = G\Delta p/\hbar \ll \theta$, that is,
\begin{equation}
 \frac{\Delta p}{\hbar} \ll \frac{\theta}{G} \simeq \frac{\tan \theta}{G} = 
  \frac{1}{G|\Re\bracket{\hat{A}}\sub{w}|}. \label{eq:27}
\end{equation}
This condition can be related to the weakness condition mentioned in
\cite{duck_sense_1989} and \cite{parks_observation_1998}.
The requirement of the weakness condition comes from the fact that
the Pancharatnam phase that is obtained by the phase jump is limited to $\pi$,
i.e. a quarter of the solid angle of the Bloch sphere.
Since the weak value is determined from the gradient of the Pancharatnam phase,
in order to obtain a large weak value, we must prepare a probe wavepacket
having a small momentum variance so that it can be confined within the linear region.

\section{Summary}
In this paper, we introduced the interferometer for particles having internal
degrees of freedom, which is a framework common to quantum eraser and weak measurement.
We first examined the phase change in quantum eraser.
It turned out that the post-selection in quantum eraser plays a role to
change the way of the phase comparison between internal states.
As a result, when we post-select the internal state, the Pancharatnam phase
appears as an additional phase shift of interference pattern.

Subsequently, we considered the weak measurement in the interferometric framework
with relating it to the quantum eraser.
We also focused on the phase change in weak measurement,
and demonstrated that the extraordinary displacement in weak measurement
is caused by the Pancharatnam phase that is obtained by post-selection.
The unbounded weak value is achieved by utilizing the
phase jump in the Pancharatnam phase.
The weakness condition can be also derived from the nonlinear property of the
Pancharatnam phase.

We hope that
our interpretation of the weak measurement, which is based on
the interferometry utilizing the Pancharatnam phase,
will enable us to comprehensively understand the weak measurement,
thereby allowing us to develop useful applications.

\ack
We thank Yutaka Shikano for interesting and inspiring
discussion.
This research was supported by the Global COE program
`Photonics and Electronics Science and Engineering' at Kyoto University.

\end{document}